\documentclass[12pt,preprint]{aastex}

\slugcomment{}

\newcommand{\cmsq}{cm$^{-2}$}
\newcommand{\kms}{km\thinspace s$^{-1}$}

\shortauthors{Taylor \& Wang}
\shorttitle{Origin of the Dust Arch in NGC~4631}

\begin{document}

\title{The Origin of the Dust Arch in the Halo of NGC~4631: An 
Expanding Superbubble?}

\author{Christopher L. Taylor}
\affil{Five College Radio Astronomy Observatory, University of Massachusetts}
\authoraddr{University of Massachusetts, Five College Radio Astronomy 
Observatory, 619B Lederle GRT, Amherst, MA 01003 }

\author{Q. Daniel Wang}
\affil{Dept. of Astronomy, University of Massachusetts}
\authoraddr{Dept. of Astronomy, University of Massachusetts, Amherst, MA 01003}

\vspace{.75in}


\begin{abstract}

We study the nature and the origin of the dust arch in the halo of the
edge-on galaxy \objectname{NGC 4631} detected by \citet{ND}.  We present 
CO observations 
made using the new On-The-Fly mapping mode with the FCRAO 14m telescope, and 
find no evidence for CO emission associated with the dust arch.  Our 
examination of previously published HI data shows that {\it if} previous 
assumptions about the dust temperature and gas/dust ratio are correct, then
there {\it must} be molecular gas associated with the arch, below our 
detection threshold.  If this is true, then the molecular mass associated
with the dust arch is between 1.5~$\times~10^8$ M$_\odot$ and 
9.7~$\times~10^8$ M$_\odot$, and likely towards the low end of the range.
A consequence of this is that the maximum allowed value for the CO-to-H$_2$ 
conversion factor is 6.5 times the Galactic value, but most likely closer to
the Galactic value.  The kinematics of the HI apparently associated with the 
dust arch reveal that the gas here is not part of an expanding shell or 
outflow, but is instead two separate features (a tidal arm and a plume of HI 
sticking out into the halo) which are seen projected together and appear as a 
shell.  Thus there is no connection between the dust ``arch'' and the hot X-ray
emitting gas that appears to surround the galaxy \citep{We01}.

\end{abstract}

\section{Introduction}

\objectname{NGC 4631} is one of the prototypical edge-on spiral galaxies 
($i \sim 86\deg$).  Such edge-on systems allow the study of the interaction 
between galaxy disks and halos and \objectname{NGC 4631} displays a number 
of phenomena associated with disk/halo interaction.  It is interacting with 
two companions, \objectname{NGC 4627} and \objectname{NGC 4656}, a fact which
may be responsible for the high level of star formation in its disk 
\citep{Re92}.  This star formation activity has lead to an outflow from 
the disk driven by the input of kinetic energy into the interstellar medium
(ISM) by massive stars via stellar winds and supernovae (e.g. \citet{ES, 
We95,HWR}).  Close to the disk \citet{R2000} has found evidence for an
outflow of molecular material in interferometric CO observations of the
central part of the galaxy.  In the disk beneath this feature are a group
of bright HII regions, a peak of radio continuum emission, and a bright
X-ray feature, all suggesting a possible starburst driven outflow.
\citet{ADB} report a possible dust outflow seen in submm continuum emission
at a distance of $\sim 1 $ kpc above the disk.  

Recently \citet{ND} have detected 1.2 mm continuum emission from cold
dust in the halo out to distances of 10 kpc. The more distant emission
appears associated with an HI tidal feature resulting from the interactions
with companions, but directly over the central region in 
\objectname{NGC 4631}, the dust emission forms an arch.  The arch encloses 
an area of enhanced X-ray emission in the halo \citep{We01}, with a 
morphology very suggestive of a shell of cool gas (associated with the 
dust), containing a hotter, X-ray emitting gas that may be part of an 
outflow from the disk.

The nearly edge-on starburst galaxy \objectname{M82} has been shown to have 
an outflow of molecular gas from the disk into the halo (\citet{TWY,SC}),
so we investigate the possibilities that either an expanding supershell or
an outflow including entrained molecular and atomic gas may be responsible 
for the dust arch in \objectname{NGC 4631}.  This galaxy has been much observed
in various CO lines recently \citep{GW94,Ye95,R2000,De01,Pe01} but these have 
all focused on the star-forming disk.  We carried out a mapping program 
with the FCRAO 14m telescope to search for CO emission from molecular gas 
in the {\it halo} of \objectname{NGC~4631}.  We also reanalyzed previously 
published HI data to search for evidence of an outflow near the dust arch. 
In Section 2 we present our observations, in Section 3 we present our results 
and discuss their implications, and in Section 4 we summarize the paper. 


\section{Observations}

The observations were obtained with the Five College Radio Astronomy 
Observatory (FCRAO) 14~m telescope over several runs in January and 
February, 2002. We observed the $^{12}$CO J = 1$\rightarrow$0 line using 
SEQUOIA, a focal plane array receiver, consisting of 16 pixels.  The 
On-The-Fly (OTF) observing mode was used, in which the telescope is scanned 
in rows over the area to be mapped while the receivers are continuously read 
out.  Using OTF mapping with SEQUOIA allowed us to achieve high sensitivity
and flat baselines over a large sky area because each position in a map is 
observed by each of the 16 pixels.  Thus we obtain a factor of 4 improvement 
relative to a single element receiver, plus any pixel-to-pixel irregularities 
are smoothed away by averaging over all 16 pixels.
We used a readout rate of 1 spectrum per second per receiver.  An area
10\arcmin~$\times~$6\arcmin .75 was mapped in this way.  One map took 
approximately 15 minutes to complete, and the map was repeated 38 times
to achieve low rms. The extragalactic filterbanks were used to obtain a 
bandwidth of 320 MHz with 5 MHz (13 \kms) channels.  System temperatures 
ranged from $\sim$ 450 to 800 K.  The pointing was checked every 2 to 4 
hours each session using SiO maser sources.

The processing of the OTF scans into a uniformly gridded final map was 
carried out using an in-house FCRAO reduction package called OTFTOOL
(see 

\anchor{http://donald.phast.umass.edu/~fcrao/library/manuals/otfmanual.html}{
{\it http:$\slash\slash$donald.phast.umass.edu$\slash\sim$fcrao$\slash$library$\slash$manuals$\slash$otfmanual.html} }

for more information regarding the implementation of OTF mapping at FCRAO).  
OTFTOOL allows the inspection, editing and regridding of the raw OTF
data. Once the map of spectra was produced, further reduction and analysis 
was done with CLASS.  A main beam efficiency of $\eta_{MB}$\,=\,0.45 was 
used used to put the spectra on the T$_{MB}$ scale.  We smoothed the data 
spatially to 60\arcsec\ to search for extended, low S/N emission.  Unless 
otherwise stated, all discussion refers to the smoothed data.  

\section{Results and Discussion}

\subsection{CO Data: The Molecular Content of the Gas in the Halo of NGC~4631}

\begin{figure*}
\plotone{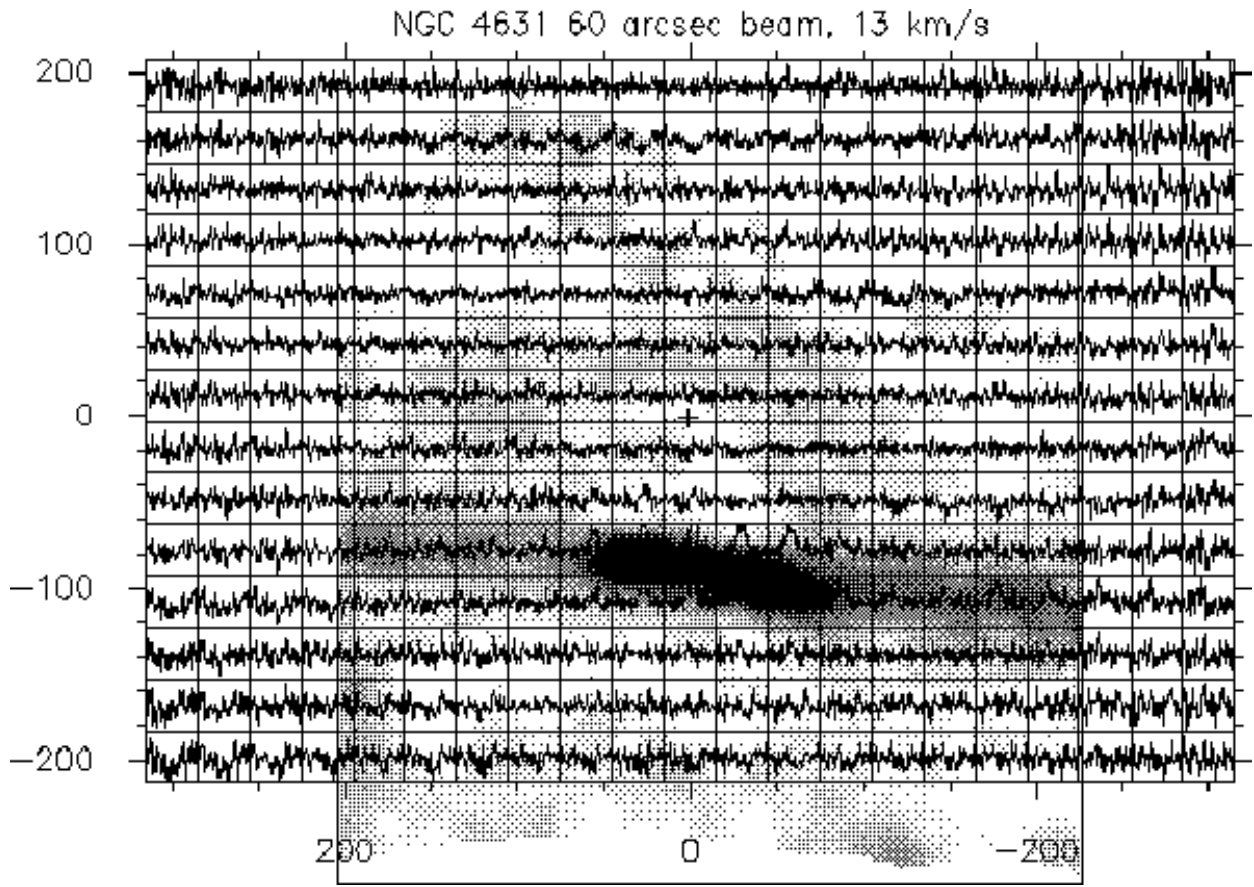}
\caption[CTaylor.fig1.ps]{Our map of CO (1-0) emission superposed over the 
central region of the dust emission map of \citet{ND}.  The CO data cover a 
10\arcmin $\times$ 6\arcmin .75 area centered north of NGC 4631.  Each 
plotted spectrum runs from 200 to 1000 \kms\ and -0.08 to 0.08 T$_{mb}$.  
The spectra shown retain the original velocity resolution of 13 \kms. 
\label{fig1} }
\end{figure*}

Figure~1 shows the spectra of our final map displayed at the original 
velocity resolution of 13 \kms\, superposed over the dust emission observed
by \citet{ND}.  CO emission from \objectname{NGC 4631} itself is clearly 
present, and corresponds quite closely with the peak of the dust emission 
at the center of the galaxy.  The total integrated intensity from the galaxy 
is 45.6 $\pm$ 4.9 K \kms, compared to 31.4 $\pm$ 1.8 K \kms\ detected by 
\citet{GW94} using the IRAM 30-m telescope.  These authors note that they 
did not map the entire galaxy and therefore consider their number a lower 
limit.  Their map has a handful of positions out to about 40\arcsec\ above 
the plane of the disk and didn't cover the dust feature of \citet{ND}.

We do not detect CO (1-0) emission coming from the region of the dust arch.
The rms typical of the spectra in this map is 24 mK.  This corresponds to
a 3$\sigma$ upper limit over 3 consecutive velocity channels of 2.9 K \kms.
If we assume a standard Galactic CO-H$_2$ conversion factor (e.g. 
2.3~$\times~10^{20}$ cm$^{-2}$/ K \kms\ from \citet{Se88}), this corresponds
to an upper limit in column density of 6.6$\times~10^{20}$ cm$^{-2}$ in
molecular gas throughout the dust arch.  The upper limit for molecular
gas mass integrated over the area of the dust arch comes to 3.4$\times~10^8$
M$_\odot$ for the commonly assumed distance of 7.5 Mpc.   We also smoothed 
the data in velocity to 39 \kms, achieving an rms of 18 mK, but still found 
no CO emission.  If we average together all the spectra in Figure~1 that
overlap the dust emission to search for extended low surface brightness CO 
emission (Figure~2), we obtain an rms of 11 mK, for a 3$\sigma$ upper 
limit of 1.3 K \kms, or 1.5$\times~10^8$ M$_\odot$.

For comparison, we note that in our observations of the halo of 
\objectname{M82} \citep{TWY} the CO emission had a strength of 36 mK at the 
same height as the dust arch in \objectname{NGC~4631}.  Thus the rms in
our map is almost, but not quite, low enough that we could detect similar
emission in \objectname{NGC~4631}. However, when we average together
all the spectra from the region of the dust arch, the sensitivity improves
and we should be able to detect a similar level of emission as is seen in
\objectname{M82}.  The fact that we don't suggests a genuine distinction
between these two edge on systems that both have substantial winds extending
into their halos.

At (0,100) in Figure~1 there appears a feature with a S/N ratio of about
3.5$\sigma$.  Although this location is above the dust arch, it does 
coincide with dust emission from the tidal feature seen in HI.  We received
additional observing time to verify this potential CO emission, integrating
for eight hours on this location in a ``staring'' mode -- {\it i.e.} not in 
a mapping mode.  After reaching an rms of 6.7 mK, we find no evidence for 
this feature in the followup data, suggesting that the peak in Figure~1 comes 
from random noise.

\begin{figure*}
\plotone{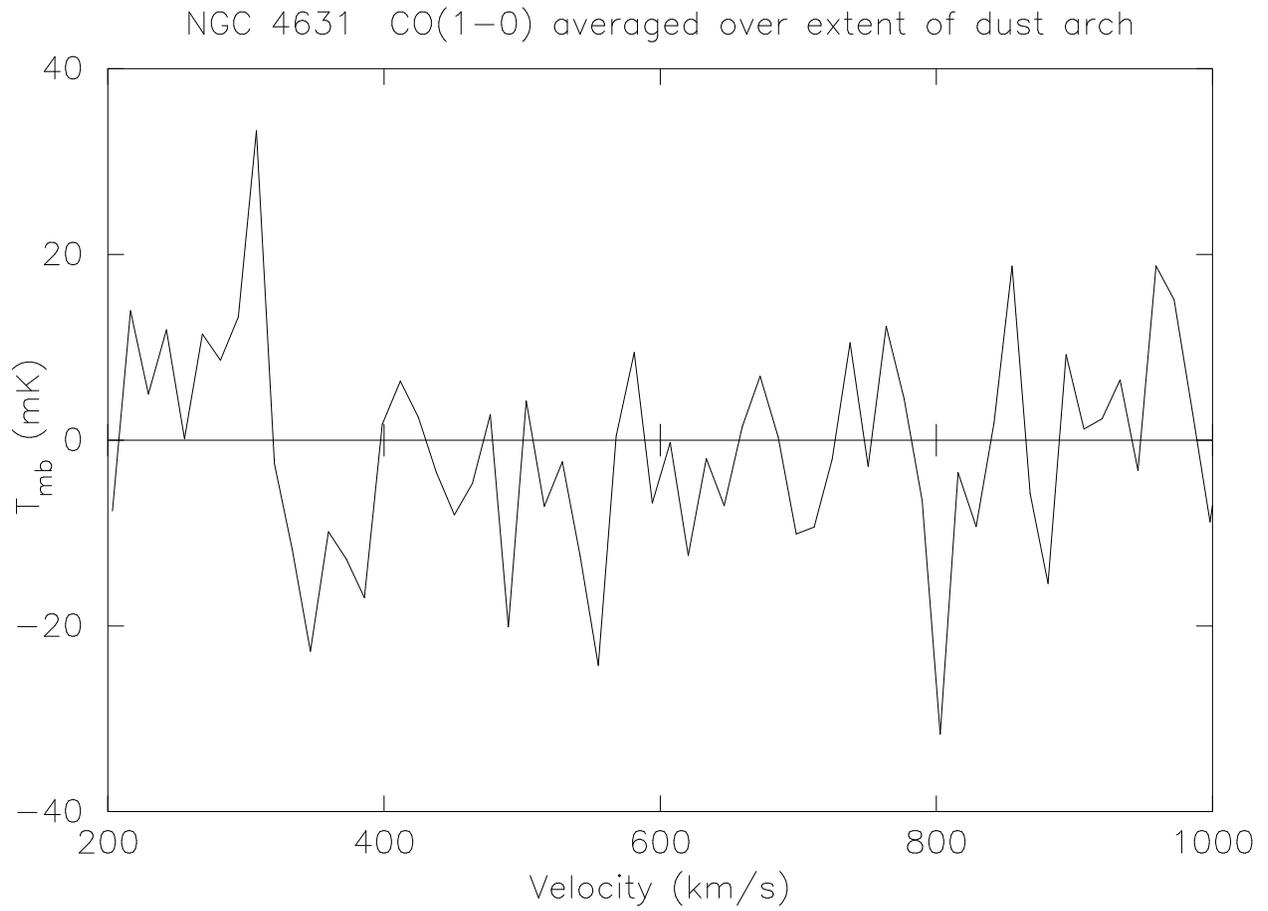}
\caption[CTaylor.fig2.ps]{The average of all spectra spatially coincident with 
with the dust arch in the halo of NGC~4631.
\label{fig2} }
\end{figure*}

\subsubsection{The Maximum Molecular Mass in the Dust Arch}

There are at least three possible explanations for our lack of a 
detection of CO emission in the region of the dust arch: 1) there 
is no molecular gas there; 2) there is molecular gas, but the associated
CO emission is below our detection threshold; 3) there is CO emission,
but it is clumped on scales much smaller than our beam and is undetectable.
We discuss each of these possibilities below.

{\bf Explanation 1:} \citet{ND} estimated the total amount of gas present 
in the dust arch based upon the dust mass and assuming a gas/dust ratio.  
We have obtained the HI data of \citet{R94} and estimated the HI associated 
with the dust arch.  The total gas mass is 1.7~$\times~10^9$ M$_\odot$, 
and the HI mass is 7.3~$\pm 0.7 \times~10^8$ M$_\odot$, {\it requiring 
$\sim$ 9.7~$\times~10^8$ M$_\odot$ of molecular gas to account for all of 
the observed dust emission.}
Thus it seems likely that explanation 1 is not correct -- we appear to require
at least some molecular gas in the dust arch region.  One caveat is that
the calculation of total mass by \citet{ND} requires the dust temperature
and a gas/dust ratio.  \citet{ND} argued that the disks of spiral galaxies
with moderate star formation activity have dust temperatures of 15 to 20 K 
and the dust in the halo is likely cooler, but this is a major source
for uncertainty.  If the temperature is 5 K cooler, then the inferred
gas mass is a factor of 2 greater, but if the temperature is 5 K warmer the
mass is 30\% less than their estimate. \citet{ND} do not say what gas/dust 
ratio they used to estimate the gas mass associated with the dust, but
likely they used a standard Galactic ratio.  It is possible to do away with
the need for molecular gas -- if the dust temperature is 30 K or hotter,
or if the gas/dust ratio is lower than Galactic by at least a factor of 2.5. 
Existing data cannot speak to either of these possibilities, while the
assumptions of \citet{ND} are plausible.

{\bf Explanation 2:} Our upper limit on molecular gas in the dust arch
is 1.5~$\times~10^8$ M$_\odot$, 6.5 times lower than the molecular gas
required to explain the dust emission, 9.7~$\times~10^8$ M$_\odot$. If
the assumptions of \citet{ND} are correct, we should see CO emission.
One way to hide CO emission when there is substantial molecular gas 
present is to increase the CO-to-H$_2$ conversion factor, such that 
a large amount of molecular gas has low CO emission.  In our calculation
of the upper limit, we assumed a typical Galactic value.  However, under
conditions of low metal abundance or strong UV radiation fields where
CO is photodissociated while the H$_2$ is largely intact, the conversion 
factor increases.  Such conditions can be found in dwarf galaxies
\citep{W95,TKS}, or in the outer regions of spiral galaxies \citep{RH94}.
If the gas in the dust arch originated from the outskirts of \objectname{NGC 
4631} (e.g. by being drawn up out of the plane due to a gravitational 
interaction with a nearby galaxy) the conversion factor in the gas could be
lower than is traditionally assumed.  These physical conditions do not
necessarily decrease the dust emission -- \citet{I97a,I97b} finds plenty 
throughout low metallicity galaxies like the \objectname{Large Magellanic 
Cloud} and \objectname{NGC 6822}.

However, if the CO-to-H$_2$ conversion factor is too high, 
then our upper limit on CO emission would correspond to too much H$_2$ mass.  
{\it This allows us to constrain the CO-to-H$_2$ conversion factor in the 
region of the dust arch.} The ratio of expected H$_2$ mass to our upper limit 
yields the maximum possible conversion factor: 

$$ { 9.7~\times~10^8 \over 1.5\times~10^8 } = 6.5 $$

The molecular mass of the disk of \objectname{NGC 4631} is $\sim 
10^9$ M$_\odot$ \citep{GW94}.  It is unlikely that a starburst driven
outflow could push half the galaxy's total molecular mass out into
the halo but leave the gas still in the disk relatively undisturbed,
so most likely 9.7~$\times~10^8$ M$_\odot$ is an overestimate for the 
halo molecular gas, and thus the CO-to-H$_2$ conversion factor is 
probably close to Galactic.  

{\bf Explanation 3:}
A further possibility to explain the large difference between our upper
limit and M$_{diff}$ is that the molecular gas is highly clumped, such that 
the emission is on scales much smaller than the observed distribution of
HI or dust.  If more that 1.5$\times~10^8$ M$_\odot$ of molecular gas
were present, and a Galactic CO-to-H$_2$ conversion factor applied, then
we would see CO emission if the molecular gas was smoothly and evenly
distributed throughout the dust arch.   However, emission could potentially 
be lost by being averaged together with noise from non-emitting areas if it
were highly clumped.  This has been shown to happen in the Local Group dwarf 
irregular galaxy \objectname{IC 10} by comparing single dish and 
interferometer CO (1-0) observations (Taylor et al., in preparation, Walter 
\& Taylor, in preparation).

By itself, the lack of CO emission from the dust arch doesn't provide us
much with information about the origin of the dust arch {\it i.e.} is it
an expanding supershell that surrounds the hot X-ray gas imaged by
\citet{We01}?

\subsection{HI Data: The Origin of the Dust Arch}

We have also reanalyzed the HI data of \citet{R94} to look for kinematic
evidence of a superwind or expanding bubble in the atomic hydrogen
spatially overlapping with the dust emission.  Figure~3 shows the 1.3 mm 
continuum dust emission with HI contours superimposed.  The HI corresponds 
well to the dust arch feature.  At the top of the dust arch, a tidal arm 
joins to the galaxy (HI spur 4 from \citet{R94}), complicating the 
interpretation.  There is clearly dust emission associated associated with 
spur 4 and it joins smoothly onto the western part of the dust arch.

\begin{figure*}
\plotone{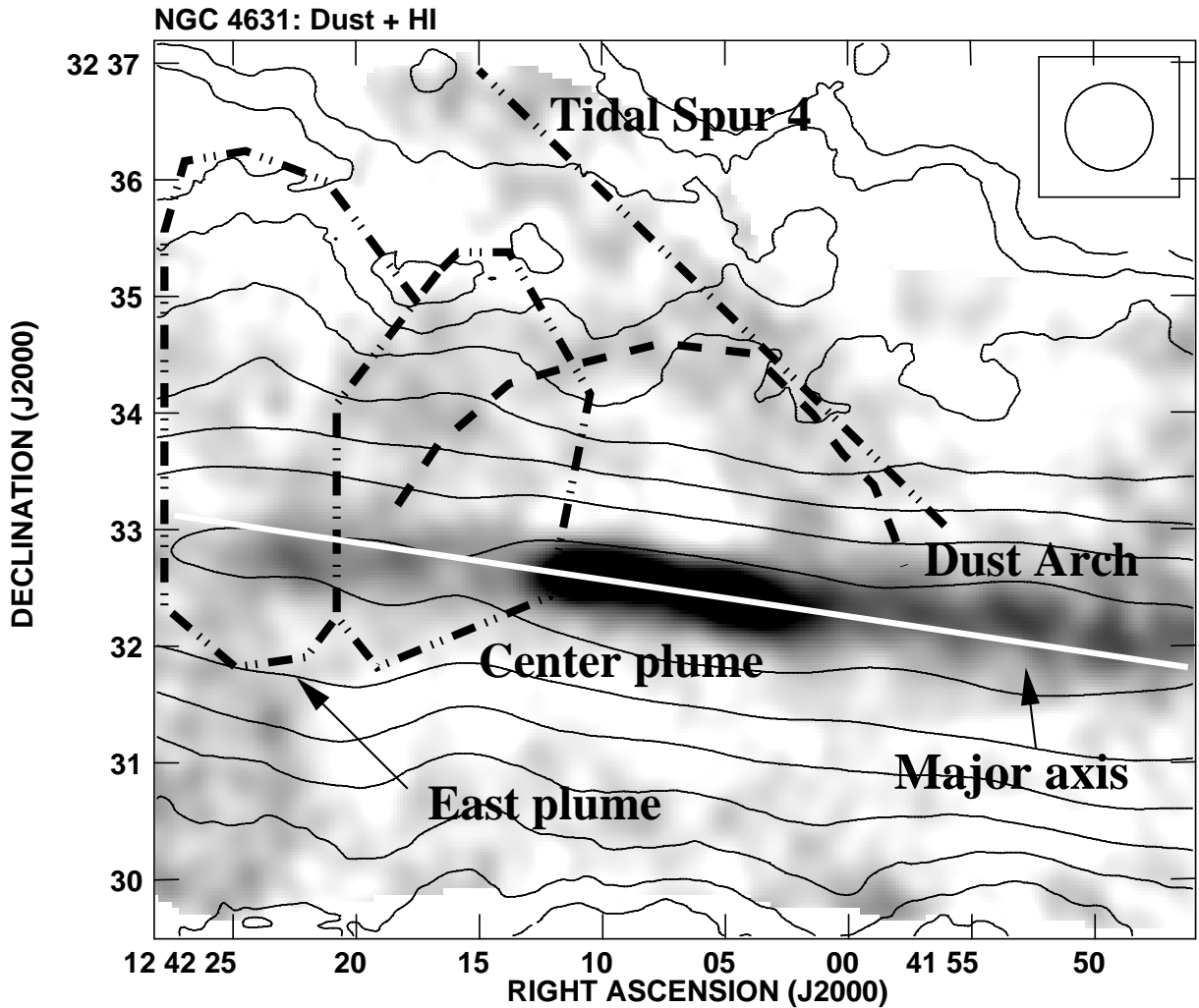}
\caption[CTaylor.fig3.ps]{Dust emission from the central region of NGC~4631 
with HI contours superposed.  The HI contours represent 1, 2, 4, 8 16, 32 and 
64 $\times 10^{20}$ \cmsq.  The map of dust emission comes from \citet{ND} and 
the HI data from \citet{R94}.  The dust arch is indicated by a dashed line.
Kinematic features from Figure~4 are indicated by dot-dash lines and are 
labeled.  The major axis is indicated by the white line.
\label{fig3} }
\end{figure*}

Because of this apparent ambiguity between the dust arch and spur 4 in the 
maps of 1.3 mm continuum emission and integrated HI intensity, we plot in 
Figure~4 position-velocity (p-v) diagrams taken parallel to the major axis 
of \objectname{NGC 4631}, at heights above the disk such that they pass 
through the region of the dust arch.  Several HI features are labeled in 
the p-v diagrams.   At velocities just above 600 \kms\ in the p-v cut
taken at 83$\arcsec$ above the disk spur 4 can be seen where it joins
the disk on the right side of the galaxy disk.  As the cuts move closer
to the plane of the galaxy, emission associated with the HI disk becomes
more prominent (going from bottom to top in Figure~4).  Closer to the 
disk, at the left side of the p-v diagrams at velocities $<$ 700 \kms\
is spur 1, visible in the HI maps of \citet{R94}.  Labeled in the p-v
diagram at 80$\arcsec$ above the disk is the ``east plume''.  This
feature is visible in Figure~3 projecting up out of the disk at 
R.A. = 12:42:26.  This eastern plume is directly above one of the HI 
supershells discovered by \citet{RvdH} and is visible in their Figure
4a.  Another plume, labeled as the ``center plume'', is most prominent
at 74$\arcsec$ above the disk.  This feature is seen in Figure~3 where
it overlaps spatially with the eastern arm of the dust arch.  There is no
HI supershell evident below the center plume, and it may tidally formed.
We find no kinematic evidence for an expanding shell of HI at the position 
of the dust arch.  In a p-v diagram this would show up as a ring shaped
region of enhanced emission.

\begin{figure*}
\plotone{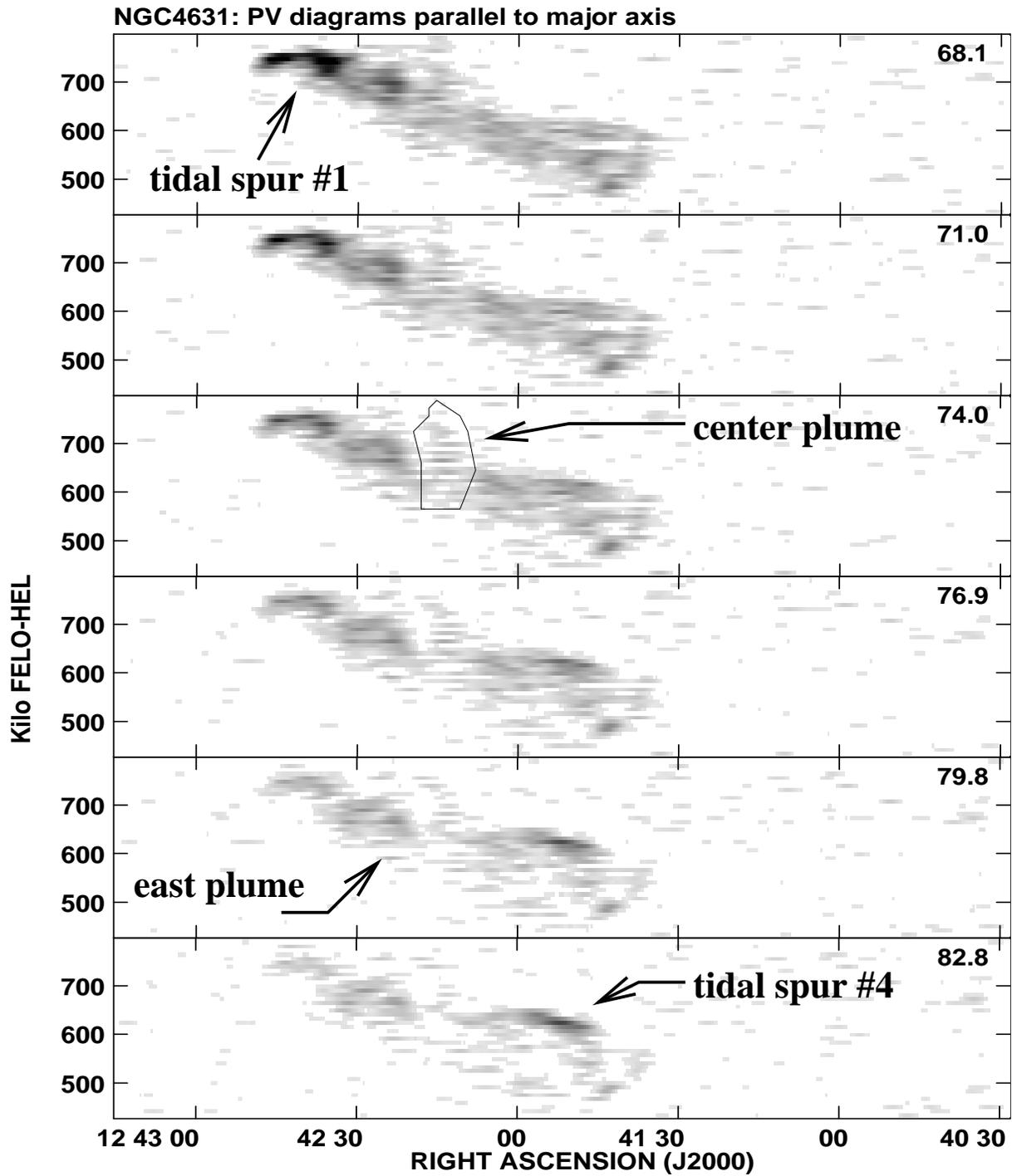}
\caption[CTaylor.fig4.ps]{Position velocity diagrams from the HI data cube, 
taken parallel to the major axis of NGC~4631, above the disk.  The cuts pass 
through the position of the dust arch and are labeled with their height above 
the major axis in arcseconds.  Features visible in Figure~3 are also labeled.
\label{fig4} }
\end{figure*}

The HI in the region of the dust arch does 
not have a single kinematic identity, but instead comes from two 
distinct features whose proximity gives the appearance of a single feature.  
The most obvious of these distinct features is the tidal arm extending 
from southwest to northeast across the halo, known as spur 4.  Clearly there 
is dust associated with spur 4, out to the extent of the region mapped by 
\citet{ND}, and joining smoothly onto the western arm of the arch.  The 
kinematics of the HI in the western arm also show that the gas here is 
part of spur 4 and not part of an expanding shell or outflow.  The HI 
spatially coincident with the eastern arm of the arch has kinematics 
distinct from the western side.  The spatial distribution of the HI here
and its kinematics is similar to that of the east plume, suggesting
that the eastern arm of the arch is a weaker HI plume that was not 
noticed by \citet{RvdH}.  These authors found two other such features
(calling them ``worms''), so it is not unlikely that there may be more.
{\it Thus the dust arch is formed from dust associated with gas that may 
only appear to form an arch in projection -- tidal spur 4 which forms the 
west of the arch and the ``cap'', and an HI plume extending up from the 
disk of \objectname{NGC~4631}.}  It is only coincidence that the apparent 
arch appears perched suggestively over the X-ray gas in the halo.

\subsection{The Dust Arch at Other Wavelengths}

\objectname{NGC~4631} is
famous for its large radio continuum halo.  It has been observed at a 
variety of resolutions at several frequencies (e.g. \citet{GH94}, 
\citet{HD90}) and there are no indications of the arch morphology being
traced in the radio continuum.  This is not to say that the radio 
continuum halo is featureless -- there are a number of radio spurs that 
extend up into the halo.  Some of these radio spurs are spatially coincident
with the HI spurs, including spur 4 that we discussed above.  \citet{HD90}
argue that the radio spurs are caused by magnetic field lines that have
been pulled up out of the disk by the gravitational interactions 
\objectname{NGC~4631} is experiencing.  The fact that the radio continuum
emission in the halo does not have an arch shape in the vicinity of the
dust arch, but does follow HI spur 4 closely suggests that the dust arch
is not a single, coherent physical structure.  If the dust arch traced 
gas that had been blown up out of the disk of \objectname{NGC~4631} by
intense star formation, we would expect the magnetic field lines to have
been carried along with it, and that should be reflected in the morphology
of the radio continuum emission.

\citet{We01} observed \objectname{NGC~4631} using ACIS on {\it Chandra},
finding that the low energy (0.3 -- 0.6 keV) diffuse X-ray emission 
appears to be contained by the dust arch.  However, they also point out
an absorption feature in this X-ray emission along the minor axis of
the galaxy, coinciding with the location of the dust arch.  We have 
identified this part of the dust arch as belonging to HI spur 4, so
an equally valid interpretation is that spur 4 is between us and the
bulk of the X-ray emission.  This is consistent with our interpretation
that the arch shape is formed by the chance superposition of spur 4 and
a kinematically distinct HI plume.

\section{Summary and Conclusions}

We used the new OTF mapping capability at the FCRAO 14m telescope to search 
for CO emission in the halo of the edge-on galaxy \objectname{NGC 4631}, in 
particular to look for molecular gas associated with the dust arch detected 
in the halo by \citet{ND}.   We also examine the HI emission in the region 
of the dust arch using data from \citet{R94}.  If previous assumptions by 
\citet{ND} regarding the dust temperature and gas/dust ratio are correct, 
we show there must be molecular gas present, because the HI alone 
wouldn't have enough dust to explain the observed millimeter continuum 
emission.  We set tight upper limits on CO emission and show that for the 
assumptions made by \citet{ND}, the total molecular gas must fall in the range:
1.5~$\times~10^8$ M$_\odot \leq $ M$_{mol} \leq$ 9.7~$\times~10^8$ M$_\odot$.
Since the upper end of that range is almost equal to the total molecular
gas in the disk, we favor an amount at the low end of the range.  The 
maximum value of the CO-to-H$_2$ conversion factor consistent with our 
range for the molecular mass is 6.5 times the standard Galactic value, but
it is probably much closer to Galactic.  We also studied the kinematics of 
the HI that is cospatial with the dust arch.  This revealed no evidence for 
an expanding shell or outflow of HI.  Instead, the arch feature is probably 
formed from two kinematically distinct features, a tidal arm that runs across 
the halo of \objectname{NGC 4631}, and a plume of HI jutting up out of the 
disk.  The proximity of these two features gives them the appearance of an 
arch.

\begin{acknowledgements}

The authors are grateful to M. Dumke for providing us with the dust emission
map and to R. Rand for providing us with the HI data cube.  We thank the 
anonymous referee for helpful comments.  The Five College 
Radio Astronomy Observatory is operated with the permission of the 
Metropolitan District Commission, Commonwealth of Massachusetts, and with 
the support of the NSF under grant AST01--00793.

\end{acknowledgements}

\vfill \eject

\end{document}